\documentclass[final]{aipproc}

\layoutstyle{6x9}

\SetInternalRegister\hbadness{8000} 

\newcommand\doingARLO[2][]{%
  \ifx\mmref\undefined #1\else #2\fi
}

\begin{document}

\title 
      [The High Redshift Mass-Metallicity Relation]
      {
The mass-metallicity
relation at  $z\sim0.7$}


\author{S. Savaglio}{
  address={Johns Hopkins University,
Baltimore, MD, USA},
  email={savaglio@pha.jhu.edu}
  thanks={This work was commissioned by the AIP}}

\iftrue
\author{K. Glazebrook}{
  address={Johns Hopkins University,
Baltimore, MD, USA},}

\author{D. Le Borgne}{
  address={University of Toronto, Toronto, ON, Canada}}

\author{S. Juneau}{
  address={Herzberg Institute of Astrophysics, National Research Council,
Victoria, BC, Canada}}

\author{R. G. Abraham}{
  address={University of Toronto, Toronto, ON, Canada}}

\author{D. Crampton}{
  address={Herzberg Institute of Astrophysics, National Research Council,
 Victoria, BC, Canada}}

\author{P. J. P. McCarthy}{
  address={Observatories of the Carnegie Institution of Washington, 
Pasadena, CA, USA}}

\author{H.--W. Chen}{
 address={Massachusetts Institute of
Technology, Cambridge, MA, USA}}

\author{R. Marzke}{address={San Francisco State University, San Francisco, 
CA, USA}}
\author{R. G. Carlberg}{address={University of Toronto, Toronto, ON, Canada}}
\author{I. J{\o}rgensen}{address={Gemini Observatory, Hilo, HI, USA}}
\author{K. Roth}{address={Gemini Observatory, Hilo, HI, USA}}
\author{I. Hook}{address={Oxford University, Oxford, England}}
\author{R. Murowinski}{address={Herzberg Institute of Astrophysics, 
National Research Council, Victoria, BC, Canada}}

\fi

\copyrightyear  {2001}

\begin{abstract}
The ISM metallicity and the stellar mass are examined in a sample of
66 galaxies at $0.4<z<1$, selected from the Gemini Deep Deep Survey
(GDDS) and the Canada-France Redshift Survey (CFRS).
We observe a mass-metallicity relation similar to that seen in
$z\sim0.1$ SDSS galaxies, but displaced towards higher masses and/or
lower metallicities. Using this sample, and a small sample of
$z\sim2.3$ LBGs, a redshift dependent mass-metallicity relation is
proposed which describes the observed results.
\end{abstract}

\date{\today}

\maketitle
\subsection{Introduction}

The mass and the metallicity of galaxies are important physical
parameters which are exhaustedly studied in the local Universe. At low
redshift ($z\sim0.1$) a convincing correlation between the stellar
mass and the metallicity in the ISM was found in a sample of 53,000
SDSS star-forming galaxies (Tremonti et al. 2004). At higher redshift,
this test was never attempted, due to the difficulty to obtain mass
(which requires NIR photometry) together with metallicity (which
requires optical/NIR spectroscopy) for a large sample of faint
objects.  As an alternative parameter for mass, the galaxy optical
luminosity can be used, although this is affected by dust
extinction. At $0.3<z<1$, Kobulnicky \& Kewley (2004) found a
correlation between luminosity and mass, with an offset towards higher
luminosities with respect to the SDSS sample, suggesting a time
evolution of the luminosity-metallicity relation.  Low metallicities
(for the observed luminosities) are also detected in some LBGs at
$z\sim2.3$ (Shapley et al. 2004).  In this work we present the first
investigation of the mass-metallicity relation at high redshift.

\begin{figure}
\includegraphics[height=.41\textheight]{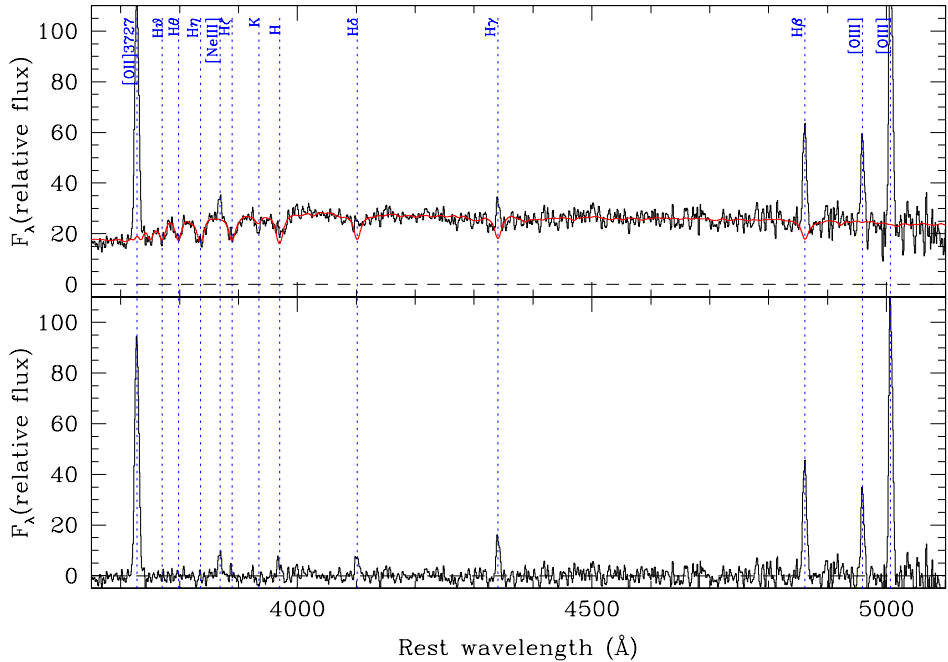}
  \caption{Composite spectrum of 26 GDDS galaxies in $0.47 < z < 0.96$
({\it upper panel}), with overplotted the ``best-fit'' stellar
population model (Bruzual \& Charlot 2003).
The {\it lower panel} shows the emission lines after the stellar
continuum subtraction.  The equivalent width Balmer absorption
correction is of the order of 3 \AA\ both for H$\beta$ and H$\gamma$,
depending on the emission line FWHM.}
\end{figure}

\subsection{The sample and the composite spectrum}

The sample selection from the GDDS (Abraham et al. 2004) is based on
the requirement that the spectra wavelength interval covers the [OII],
[OIII] and H$\beta$ emission lines (30 galaxies in $0.4 < z < 1.00$).
To measure emission line fluxes, the continuum was estimated from two
small regions before and after the line. In those cases where the
[OIII]$\lambda4958$ emission line is barely detected, the flux is
assumed to be 0.34 times lower than the [OIII]$\lambda5007$ line flux
(as set by atomic parameters).

The stellar Balmer absorption correction in the individual spectra is
done by using the mean Balmer absorption in the composite
spectrum. This is estimated by fitting the stellar continuum with
Bruzual \& Charlot (2003) stellar population synthesis models.  The
composite spectrum is also used to derive the dust extinction, via
the Balmer decrement measurement, which gives an optical extinction
$A_V=2.2\pm0.3$ (for a gas temperature $T=10,000$ K, and a Milky Way
extinction law). For 1/3 of the galaxies, the H$\gamma$ emission line
is detected, and the Balmer decrement estimated directly.

\begin{figure}
\includegraphics[height=.34\textheight]{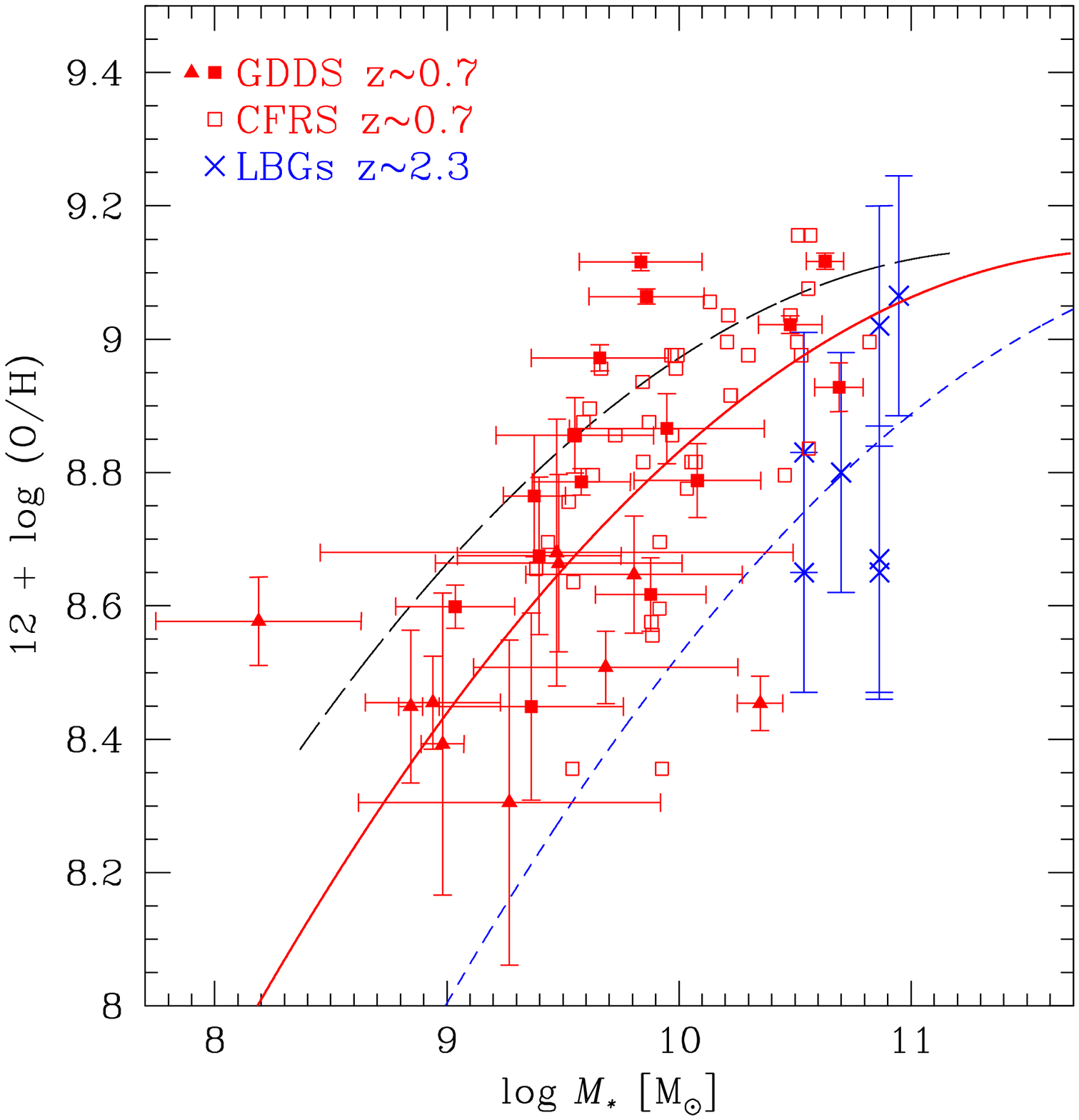}
\includegraphics[height=.34\textheight]{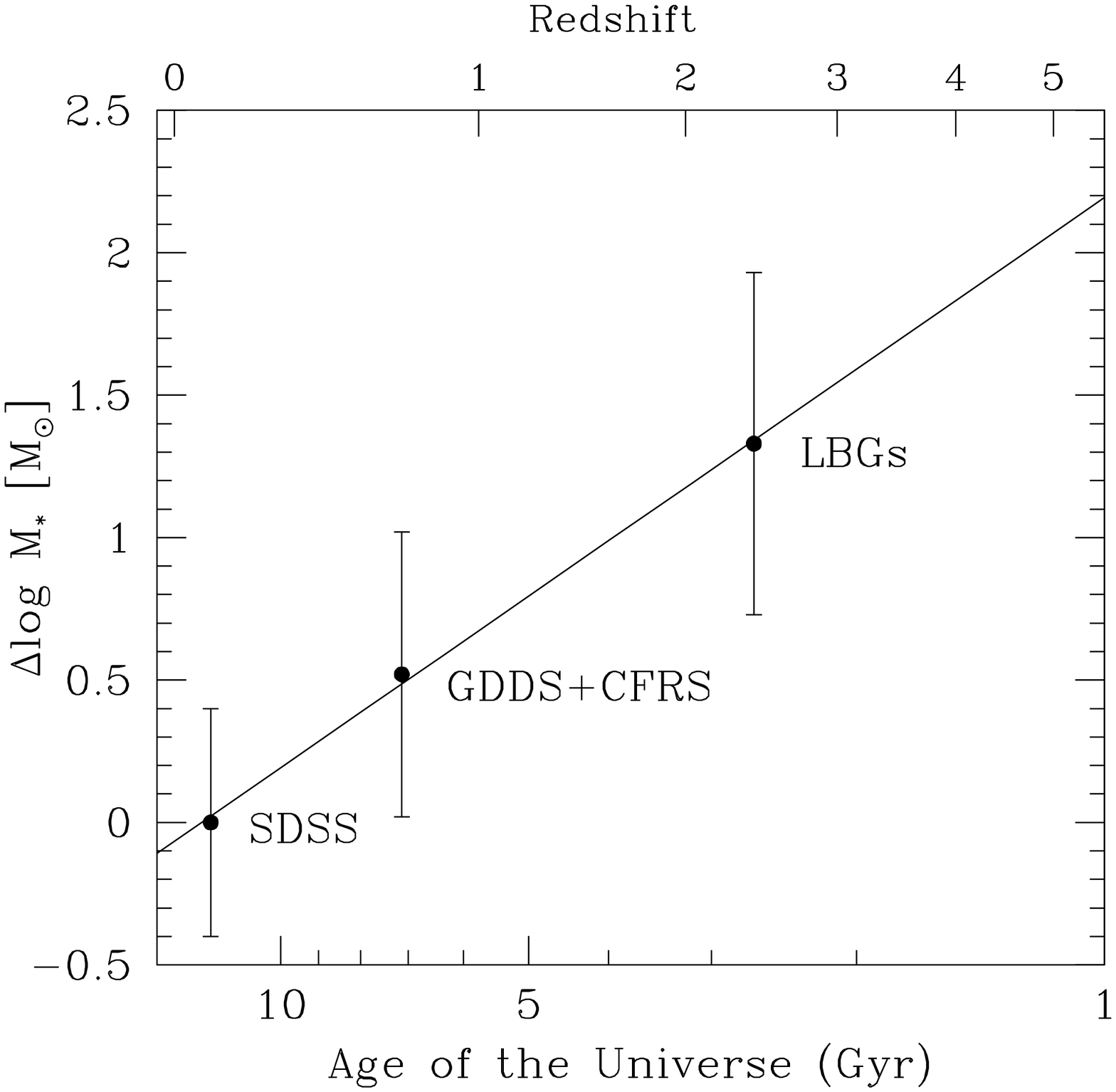}
\caption{{\it Left panel}: Mass vs. metallicity for $z\sim0.7$
galaxies in the GDDS (filled squares: $K<20.6$; filled triangles:
$K>20.6$) and CFRS (open squares), and for $2.1<z<2.4$ LBGs
(crosses). LBG metallicities are taken from Shapley et al. (2004)
after a correction in mass and metallicity which takes into account
the different IMF and metallicity calibrator used.  The left
long-dashed curve is the best fit polynomial fit to 53,000 SDSS
$z\sim0.1$ galaxies (Tremonti et al. 2004). The solid and short-dashed
lines are the same curve shifted to the right to match the GDDS+CFRS
and LBG distributions. {\it Right panel}: The shift in mass applied to
the SDSS mass-metallicity relation to match the GDDS+CFRS and LBG
distributions. Error bars indicate the $\sim1\sigma$ mass dispersion
around the mass-metallicity relation in the three samples. }
\end{figure}

\subsection{The Mass-Metallicity Relation at $z\sim0.7$}

Dust extinction and Balmer absorption corrected emission line fluxes
in the GDDS galaxies were used to estimate ISM metallicities via the
$R_{23}$ parameter (Pagel et al. 1979). The particular
metallicity:$R_{23}$ relation used is the one recently provided by
Kobulnicky \& Kewley (2004). Metallicities for the CFRS sample are
those derived by Lilly et al. (2003), after a small correction of
$+0.06\ dex$ to account for the different metallicity:$R_{23}$
relation used.  Stellar masses of GDDS and CFRS galaxies are derived
when the $K$ or $z'$ band magnitudes are measured, using the method
described in Glazebrook et al. (2004).  The left panel of Figure 2
shows metallicities vs. mass for the total sample of 25+41
galaxies. The low redshift mass-metallicity relation of the SDSS
galaxies is shown as a long-dashed line, and the same relation
displaced to the right by $\Delta \log M_\star=0.54$ as a solid
line. This displacement is interpreted as due to a redshift evolution
of the mass-metallicity relation: galaxies at high redshift tend to
have higher masses (or lower metallicities) with respect to galaxies
with the same metallicities (or same masses) in the local
Universe. This conclusion is not disproved by the results for 7 LBGs
at $z\sim2.3$ (Shapley et al. 2004), which have, for the measure
metallicities, higher stellar masses (crosses in the left panel of
Figure 2). Note that LBGs have been shifted to higher metallicities
and lower masses, to correct for the different metallicity calibrator
and IMF used. The right panel of Figure 2 shows the displacements in
mass in the mass-metallicity relation as a function of the age of the
Universe, for the three samples at redshifts $z\sim0.1, 0.7$ and 2.3.

Tremonti et al. (2004) provide a mass-metallicity polynomial fit of the form:

\begin{equation}
12 + \log {\rm (O/H)} = -1.492+1.874y-0.08026y^2
\end{equation}

\noindent
where $y \equiv \log M_\star$. If we use the fit in the right panel of
Figure 2, the parameter $y$ can be expressed by:

\begin{equation}
y=\log M_\star+2.003 \log t_H-2.194
\end{equation}

\noindent
and Eqs.~1 and 2 give a general mass-metallicity relation that is also
function of the Hubble time $t_H$ (expressed in Gyr). These two
relations together are shown in Figure 3, where metallicity is
displayed as a function of the stellar mass (for fixed redshifts, left
panel) or as a function of the Hubble time (for fixed stellar masses,
right panel).

This empirical model, which describes the joint redshift evolution of
the stellar mass and metallicities, is a very exciting result that is
certainly opening a new view towards the cosmic metal production, both
from the theoretical and the observational point of view. A galaxy
sample in a restricted mass interval, can only provide a partial 
understanding of this problem.  In the near future we will adopt
simple theoretical methods, which use different approaches for the
IMF and SN feedback, to model mass and metallicity of galaxies from
their birth to the present.

\begin{figure}
\includegraphics[height=.4\textheight]{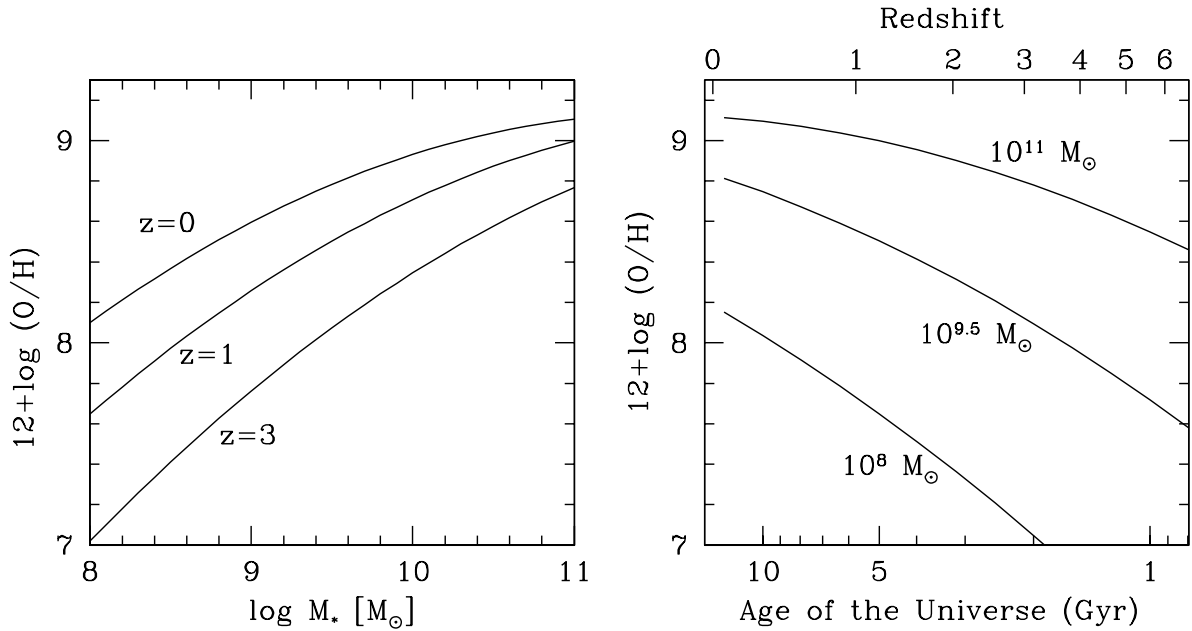}
\caption{Metallicity as a function of stellar mass, for constant
redshifts ({\it left panel}) or as a function of Hubble time for constant
stellar masses ({\it right panel}). These relations are obtained by
extrapolating at higher redshifts the Tremonti et al.'s relation
(obtained for the SDSS galaxies).}
\end{figure}



\end{document}